\documentclass[prl,twocolumn,10pt]{revtex4}
\usepackage{amsmath}
\usepackage{dcolumn}
\usepackage[cp1251]{inputenc}
\usepackage[english]{babel}
\usepackage{epsfig}
\usepackage{amssymb}

\newcommand{\sgn}{{\rm sgn}}

\begin{document}

\title{Atom lens without chromatic aberrations}
\author{Maxim A. Efremov,$^{1,2}$ Polina V. Mironova,$^1$ Wolfgang P. Schleich$^1$}
\affiliation{$^1$Institut f\"ur Quantenphysik and Center for
Integrated Quantum Science and Technology ($\it IQ^{ST}$),
Universit\"at Ulm, 89081 Ulm, Germany \\
$^2$A.M. Prokhorov General Physics Institute, Russian Academy of
Sciences, 119991 Moscow, Russia}

\email{max.efremov@gmail.com}

\date{\today}

\begin{abstract}
We propose a lens for atoms with reduced chromatic aberrations and calculate its focal length and spot size.
In our scheme a two-level atom interacts with
a near-resonant standing light wave formed by two running waves
of slightly different wave vectors, and a far-detuned running wave propagating perpendicularly to
the standing wave. We show that within the Raman-Nath approximation and for
an adiabatically slow atom-light interaction, the phase acquired by the atom
is independent of the incident atomic velocity.

\end{abstract}
\maketitle

A crucial element of the tool box for atom optics \cite{Pritchard} is a lens to focus atom waves.
Such an atom lens plays a crucial role  
in the realm of atom lithography \cite{Oberthaler-Pfau-Balykin}, which is important nowadays for a multitude of technological applications.
For this reason, many theoretical suggestions \cite{Atom-focusing} and their realizations in experiments \cite{Atom-focusing-experiments} 
have been made using laser fields. 
However, most of these realizations suffer from
chromatic aberrations. In the present paper we propose a lens,
that is free of this type of aberration by using a special combination of light waves.
Our lens is the atom-optics analog of a conventional achromatic lens \cite{MB-EW}.

We start our analysis by recalling the key features of a conventional thin lens \cite{Oberthaler-Pfau-Balykin,Sch} where
a two-level atom interacts with a standing light field detuned by $\Delta$ giving rise to the Rabi frequency $\Omega_0$.
This interaction creates the optical potential $U_{\rm opt}(x)\equiv(\hbar\Omega_0^2/\Delta)\sin^2(k_x x)$
for the motion along the $x$-axis, which we treat quantum-mechanically.
In contrast, the velocity $v_y$ of the atom in the direction
of the $y$-axis is large and remains almost constant during the
scattering process. For this reason we consider this motion
classically, which allows us to set $y\equiv v_y t$. 

Moreover, due to the small interaction time $w_0/v_y$ determined by the waist $w_0$ of 
the standing wave and the longitudinal velocity $v_y$, and the large detuning $\Delta$, 
we neglect the spontaneous emission, provided that
\begin{equation}
 \label{spontaneous emission}
    \Gamma w_e\left(\frac{w_0}{v_y}\right)\lesssim 1,
\end{equation}
where $\Gamma$ and $w_e$ are the spontaneous emission rate and
the occupation probability of the excited state, respectively.
In the case of $|\Delta|>\Omega_0$, the maximum value of
the population probability is $w_e\sim(\Omega_0/\Delta)^2$.  

In the Raman-Nath approximation \cite{Yakovlev} for the transverse center-of-mass motion of the atom 
the displacement of the atom along the $x$-axis caused by atom-field interaction
is small compared to $1/k_x$, corresponding to
\begin{equation}
 \label{RN}
   \omega_{r}\,\Omega_0\,\frac{w_0^2}{v_y^2}\ll 1,
\end{equation}
where $ \omega_{r}\equiv\hbar k_x^2/(2M)$ denotes the recoil frequency.

Within of these approximations we imprint the  phase
\begin{equation}
 \label{phase-optical potential}
    \phi(x)\sim -\frac{U_{\rm opt}(x)}{\hbar}\frac{w_0}{v_y}=-\frac{\Omega_0^2}{\Delta}\frac{w_0}{v_y}\sin^2(k_x x)
\end{equation}
onto the wave function of the center-of mass motion of the atom in the ground state, which for $k_x|x|\ll 1$
is a quadratic function of $x$, that is
\begin{equation}
 \label{phase-node}
    \phi(x)\approx-\frac{\Omega_0^2}{\Delta}\frac{w_0}{v_y}k_x^2 x^2.
\end{equation}

This quadratic variation is the origin of a thin lens \cite{Sch,Yakovlev} with the focal length
\begin{equation}
 \label{focus-optical potential}
 {\mathcal F}_0=\frac{Mv_y^2}{2\hbar\Omega_0}\frac{\Delta}{k_x^2 w_0\Omega_0}\equiv \kappa v_y^2\Delta.
\end{equation}

The spot size ${\mathcal S}\equiv\alpha_0{\mathcal F}$ is determined by the focal length ${\mathcal F}$
and the angular divergence $\alpha_0\equiv\delta v/v_y$ of the atomic beam,
with $\delta v$ being the uncertainty of the transverse atomic velocity.
For a Gaussian wave packet of width $\delta x$ the uncertainty $\delta v=\hbar/(M\delta x)$
gives rise to the angular divergence $\alpha_0=\hbar/(Mv_y\delta x)$ and the spot size
\begin{equation}
 \label{size-optical potential}
    {\mathcal S}_0=\frac{v_y\Delta}{2k_x^2\delta x w_0\Omega_0^2}\,.
\end{equation}


\begin{figure}
\includegraphics[width=0.45\textwidth]{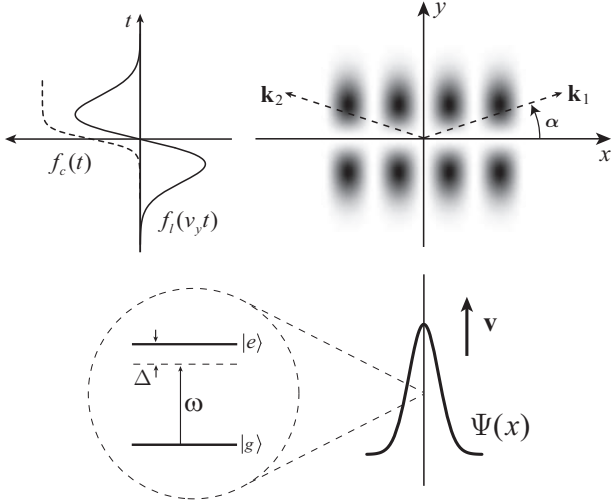}
\caption{\label{fig_setup} Scattering of the wave packet
$\Psi=\Psi(x)$ of a two-level atom by a combination of a standing electromagnetic
field formed by two propagating waves of wave vectors $\mathbf{k}_1$
and $\mathbf{k}_2$ and a traveling wave propagating orthogonal to the $x-y$-plane
serving as a control field.
As the atom propagates along the $y$-axis with the velocity $\mathbf{v}=v_y\,{\rm{\mathbf{e}}}_y$,
the envelope of the two running waves translates according to the relation $y=v_y t$
into the time-dependent function $f_l=f_l(v_y t)$.
During the atom-field interaction
the effective detuning of the field frequencies from the atomic transition
changes its sign due to the control field with an envelope $f_c(t)$
in the shape of a ''top-hat``.}
\end{figure}


According to Eqs. (\ref{focus-optical potential}) and (\ref{size-optical potential})
in a thin conventional lens both the focal length ${\mathcal F}_0$
and the spot size ${\mathcal S}_0$  depend on the atomic velocity $v_y$,
namely ${\mathcal F}_0\propto v_y^2$ and ${\mathcal S}_0\propto v_y$, resulting in large chromatic aberrations.
These scaling laws serve as our motivation to engineer a phase element for atoms and, in particular,
a lens with reduced chromatic aberrations.
Our suggestion relies on the interaction of a two-level atom with a near-resonant standing wave providing us with
the optical potential inducing the focusing, and a far-detuned traveling light wave removing the achromatic aberrations.

We create the lens field by the superposition
\begin{equation}
 \label{stwE}
    \mathbf{E}_{s}(t,\mathbf{r})=\mathbf{E}_l(\mathbf{r})e^{-i\omega t}
    \left(e^{i{\bf k}_2{\bf r}}-e^{i{\bf k}_1{\bf r}}\right)+c.c.
\end{equation}
of two traveling waves of wave vectors
$\mathbf{k}_1\equiv (k_x,k_y)\equiv(k\cos\alpha,k\sin\alpha)$ and
$\mathbf{k}_2\equiv (-k_x,k_y)$,  which form an
angle $\alpha$ relative to the $x$-axis shown in Fig. \ref{fig_setup}.
Here $\mathbf{E}_l(\mathbf{r})$ describes the position-dependent
real-valued amplitude of the waves, which, throughout the article,
is assumed to be of the form of
the ${\rm TEM}_{01}$ Hermite-Gauss mode with a node along the $y$-axis.
The frequency $\omega$ is detuned from the frequency
of the atomic transition between the ground $|g\rangle$ and excited $|e\rangle$ states
of the corresponding energies $E_g\equiv\hbar\omega_g$ and $E_e\equiv\hbar\omega_e$ by an amount
$\Delta\equiv\omega-\omega_e+\omega_g$ as shown in Fig. \ref{fig_setup}.

A running control wave
\begin{equation}
 \label{E_1}
    \mathbf{E}_{r}(t,\mathbf{r})=\mathbf{E}_{c}(x,y)e^{i(k_cz-\omega_c t)}+c.c.
\end{equation}
with the position-dependent amplitude $\mathbf{E}_{c}(x,y)$ and the shape of a ``top-hat``
propagates along the $z$-axis perpendicular to the $xy$-plane.
The frequency $\omega_c$ is far-detuned by $\Delta_c\equiv\omega_c-\omega_{e'}+\omega_e$
from the atomic transition between the exited state $|e\rangle$ and some other state $|e'\rangle$.
We suppose that the control field is weak enough to be considered perturbatively,
resulting in the Stark shift
$\Delta E_e=|\tilde{\boldsymbol{\wp}}{\bf E}_c|^2/(\hbar\Delta_c)$
of the atomic exited state $|e\rangle$,
where $\tilde{\boldsymbol{\wp}}\equiv\langle e|\mathbf{d}|e'\rangle$ is the dipole matrix element.

The time evolution of the state-vector
\begin{equation}
 \label{wfC}
    |\Psi(t)\rangle=a_e(t;{\bf r})e^{-i\omega_e t}|e\rangle+a_g(t;{\bf r})e^{-i\omega_g t}|g\rangle
\end{equation}
follows from the Schr\"odinger equation. Indeed, within the rotating-wave approximation
the time-dependent amplitudes $a_g$ and $a_e$, which depend on the position ${\bf r}$ of the atom
as a parameter, obey the system of equations
\begin{equation}
 \label{ampeqn}
    i\hbar\frac{d}{dt}{{a}_e\choose {a}_g}=\hat{H}{a_e\choose a_g}.
\end{equation}

The Hamiltonian
\begin{equation}
 \label{amp}
    \hat{H}\equiv
    \begin{pmatrix} \Delta E_e & \;\;\; V_l^*e^{-i\Delta t} \\
      V_l e^{i\Delta t} & \;\;\;0
    \end{pmatrix}
\end{equation}
contains the complex-valued coupling matrix elements
\begin{equation}
 \label{V0}
    V_l({\bf r})=2\boldsymbol{\wp}{\bf E}_l(x,y)e^{-i k_y y}\sin(k_x x),
\end{equation}
with $\boldsymbol{\wp}\equiv\langle g|\mathbf{d}|e\rangle$ being the dipole matrix element.

We assume that the $x$- and $y$-dependence of ${\bf E}_l$ and ${\bf E}_c$ can be
separated and, since the atomic motion along the $y$-axis is treated classically, $y=v_yt$,
we find the form ${\bf E}_l(x,y)\equiv\boldsymbol{\mathcal E}_l(x)f_l(y)=\boldsymbol{\mathcal E}_l(x)f_l(v_y t)$ and
${\bf E}_c(x,v_yt)=\boldsymbol{\mathcal E}_c(x)f_c(t)$ for the electric field amplitudes.
Here the envelope function
\begin{equation}
 \label{f(t)}
    f_l(y)\equiv\frac{\sqrt{2}}{\pi^{1/4}}\frac{y}{w_0}\exp\left(-\frac{y^2}{2w_0^2}\right)
\end{equation}
of the standing wave results from the ${\rm TEM}_{01}$ Hermite-Gauss mode along the $y$-axis
and satisfies the normalization conditions
\begin{equation}
 \label{norm}
    \int\limits_{-\infty}^{\infty}dy f_l^2(y)=w_0.
\end{equation}

In contrast, the envelope
\begin{equation}
 \label{f1(t)}
    f_c(t)\equiv\theta(t)=\left\{
      \begin{array}{ll} 1,\;\;\;t\geq 0\\ 0,\;\;\;t<0
      \end{array}\right.
\end{equation}
of the control field has a ''top-hat'' profile, that is a step-wise dependence
as expressed by the Heaviside function $\theta(t)$ \cite{remark}.

Moreover, the detunings $\Delta$ and $\Delta_c$ are assumed to have the same sign and
we can then set the amplitude of the Rabi frequency
$\Omega_c\equiv|\tilde{\boldsymbol{\wp}}\boldsymbol{\mathcal E}_c|/\hbar$
of the control field to $\Omega_c=\sqrt{2\Delta\Delta_c}$.
As a result, the Stark shift $\Delta E_e$ induced by the control light field is given by
$\Delta E_e=2\hbar\Delta f_c^2(t)$.

The Hamiltonian Eq. (\ref{amp}) takes finally the form
\begin{equation}
 \label{amp-final}
    \hat{H}\cong
    \begin{pmatrix}
      2\hbar\Delta f_c^2(t) & \hbar\Omega(x)f_l(v_y t)e^{-i(\Delta-\omega_\alpha)t}\;\\
      \hbar\Omega(x)f_l(v_y t)e^{i(\Delta-\omega_\alpha)t} & 0
    \end{pmatrix},
\end{equation}
where
\begin{equation}
 \label{Doppler}
    \omega_\alpha\equiv k_y v_y
\end{equation}
and
\begin{equation}
 \label{Omega}
    \Omega(x)\equiv(2|\boldsymbol{\wp}\boldsymbol{\mathcal E}_l|/\hbar)\sin(k_x x)\equiv
    \Omega_0\sin(k_x x)
\end{equation}
are the velocity-dependent Doppler and position-dependent Rabi frequencies, respectively.

We now solve the Schr\"odinger equation (\ref{ampeqn}) with the Hamiltonian Eq. (\ref{amp-final})
in the case of an adiabatically slow atom-field interaction.
For this purpose we substitute the second equation of the system (\ref{ampeqn})
for the amplitude $a_g$ into the first one for $a_e$ and get the second order differential equation
\begin{equation}
 \label{equ}
    \frac{d^2}{dt^2}a_g-\left(i\tilde{\Delta}+
    \frac{1}{f_l}\frac{df_l}{dt}\right)\frac{d}{dt}a_g+\Omega^2f_l^2 a_g=0
\end{equation}
with the initial conditions
\begin{equation}
 \label{initial}
    a_g(t_0)=1,\;\;\;\frac{da_g}{dt}\Big|_{t_0}=0
\end{equation}
at time $t_0$.
Here we have introduced the time-dependent effective detuning
\begin{equation}
 \label{delta-eff}
    \tilde{\Delta}(t)\equiv\Delta-\omega_\alpha-2\Delta f_c^2(t)=
      \Delta-\omega_\alpha-2\Delta \theta(t).
\end{equation}

In the case of a slowly varying envelope $f_l(t)$ with
\begin{equation}
    \Big|\frac{1}{f_l}\frac{df_l}{dt}\Big|\ll|\tilde{\Delta}|,\;\;\;
    {\text{or}}\;\;\;|\tilde{\Delta}|\,\frac{w_0}{v_y}\gg 1,
\end{equation}
we can neglect its time derivative in the second term of Eq. (\ref{equ}) and arrive at
the approximate equation
\begin{equation}
 \label{equ-approx}
    \frac{d^2}{dt^2}a_g-i\tilde{\Delta}\frac{d}{dt}a_g+\Omega^2f_l^2 a_g\simeq 0.
\end{equation}

For each time interval, that is for $-\infty<t\leq 0$ and $0\leq t<\infty$,
when the detuning $\tilde{\Delta}$ is constant, the solution of Eq. (\ref{equ-approx})
with the initial conditions Eq. (\ref{initial}) reads
\begin{equation}
 \label{A-solution}
  a_g(t)=\frac{1}{2}\left[1-\frac{\tilde{\Delta}(t_0)}{\lambda(t_0)}\right]e^{i\phi_{+}(t)}+
  \frac{1}{2}\left[1+\frac{\tilde{\Delta}(t_0)}{\lambda(t_0)}\right]e^{i\phi_{-}(t)}
\end{equation}
with
\begin{equation}
 \label{Phi}
    \phi_{\pm}(t)=\frac{1}{2}\int_{t_0}^{t}dt'\left[\pm\lambda(t')+\tilde{\Delta}(t')\right]
\end{equation}
and
\begin{equation}
 \label{lambda}
    \lambda(t)\equiv\sqrt{\tilde{\Delta}^2(t)+4\Omega^2 f_l^2(v_y t)}\,.
\end{equation}

For the two time intervals $-\infty<t \leq 0$ and $0\leq t<\infty$ the initial time $t_0$ corresponds to 
$t_0=-\infty$ and $t_0=0$, respectively, where the envelope $f_l(v_y t)$ vanishes. 
From Eq. (\ref{lambda}) we find $\lambda(t_0)=|\tilde{\Delta}(t_0)|$ and 
\begin{equation}
 \label{initial-coefficient}
    \frac{1}{2}\left[1\pm\frac{\tilde{\Delta}(t_0)}{\lambda(t_0)}\right]=
      \frac{1}{2}\left[1\pm\frac{\tilde{\Delta}(t_0)}{|\tilde{\Delta}(t_0)|}\right]=
      \theta\left[\pm\tilde{\Delta}(t_0)\right].
\end{equation}

With the definitions Eqs. (\ref{delta-eff}), (\ref{Phi}) and (\ref{lambda}) of $\tilde{\Delta}$, 
$\phi_{\pm}$ and $\lambda$ together with the explicit form Eq. (\ref{initial-coefficient}) for coefficients 
we can cast Eq. (\ref{A-solution}) into the
compact form $a_g\equiv \exp(i\varphi_g)$, where the phase 
\begin{equation}
 \label{total-phase}
  \varphi_g(t;x)=\frac{1}{2}\int\limits_{-\infty}^{t}dt'
  \left[\tilde{\Delta}-\sgn(\tilde{\Delta})\sqrt{\tilde{\Delta}^2+4\Omega^2(x)f_l^2(v_y t')}\,\right]
\end{equation}
depends on the transverse coordinate $x$ of the atom. Since we are interesting in
engineering a lens for matter waves, we can ignore the phase
$-\omega_g t$  in the Schr\"odinger picture, which is independent of
$x$, and the total phase $\Phi_g$ acquired by the atom during its
interaction with the two light fields is the sum
\begin{equation}
 \label{total-phase-Phi}
 \Phi_g\equiv\varphi_g(t\rightarrow \infty; x)=\phi_g(x; \Delta-\omega_\alpha)+\phi_g(x; -\Delta-\omega_\alpha)
\end{equation}
of the two contributions determined by the time intervals $-\infty<t\leq 0$ and $0\leq t<\infty$,
where
\begin{equation}
 \label{total-phase 0}
  \phi_g(x;\delta)\equiv \frac{\sgn(\delta)}{2}\int\limits_{0}^{\infty}dt
  \left[|\delta|-\sqrt{\delta^2+4\Omega^2(x)f_l^2(v_y t)}\,\right].
\end{equation}

In the case of $\omega_{\alpha}\ll|\Delta|$,
the total phase $\Phi_g$ given by Eq. (\ref{total-phase-Phi}) reduces to
\begin{equation}
 \label{total-phase-sum}
 \Phi_g(x)=\omega_{\alpha}
  \int\limits_{0}^{\infty}dt\left[\frac{|\Delta|}{\sqrt{\Delta^2+4\Omega^2(x)f_l^2(v_y t)}}-1\right].
\end{equation}

When we introduce the integration variable $y\equiv v_y t$, and
recall the definitions Eqs. (\ref{Doppler}) and (\ref{Omega}) of $\omega_\alpha$ and $\Omega$
we arrive at
\begin{equation}
 \label{total-phase-sum-result}
  \Phi_g(x)=k_y\int\limits_{0}^{\infty}dy
  \left\{\left[1+4\frac{\Omega_0^2}{\Delta^2}\sin^2(k_{x}x)f_l^2(y)\right]^{-\frac{1}{2}}-1\right\}.
\end{equation}

We emphasize that $\Phi_g$ is proportional to $k_y$, which is a consequence of
the non-collinearity of the two wave vectors ${\bf k}_1$ and ${\bf k}_2$.
Moreover, $\Phi_g$ is independent of $v_y$.
Hence, the combination of the lens field and the control wave acting on the atom creates an achromatic phase element.

We now use this phase element to construct a lens with reduced chromatic aberrations.
For this purpose we consider a position of the atom close to a node of the standing wave,
which allows us to expand the square root in Eq. (\ref{total-phase-sum-result}), and we arrive at
\begin{equation}
 \label{phase-optical potential-new}
    \Phi_g(x)\approx \frac{k_yv_y}{\Delta}\left(-\frac{\Omega_0^2}{\Delta}\frac{w_0}{v_y}k_x^2 x^2\right)
     =\frac{k_yv_y}{\Delta}\,\phi(x).
\end{equation}
Here we have recalled the normalization condition Eq. (\ref{norm})
for the profile function $f_l$ and the form Eq. (\ref{phase-node})
of the phase $\phi$ induced by a regular optical potential.

Due to the control field $\Phi_g$ is the product of the phase $\phi$
corresponding to a conventional optical potential and the ratio $(k_yv_y)/\Delta$.
Hence, the focal length and the spot size of our lens read
\begin{equation}
 \label{focus length}
    {\mathcal F}=\frac{\Delta}{k_yv_y}\,{\mathcal F}_0\;\;\;{\rm and}\;\;\;
    {\mathcal S}=\frac{\Delta}{k_yv_y}\,{\mathcal S}_0.
\end{equation}

Since according to Eqs. (\ref{focus-optical potential}) and (\ref{size-optical potential})
${\mathcal F}_0$ and ${\mathcal S}_0$ depend quadratically and linearly on
$v_y$, in our lens the focal length ${\mathcal F}$
is proportional to $v_y$ and the spot size ${\mathcal S}$ is independent of it.
This scaling implies a reduction of the chromatic aberrations in comparison with
the conventional technique of focusing atoms. Moreover, in our lens both the focal length
${\mathcal F}$ and the spot size ${\mathcal S}$ are larger by a factor $|\Delta|/(k_yv_y)$
than those of the conventional lens.

The reduced chromatic aberrations are due to the symmetry of the ${\rm TEM}_{01}$ Hermite-Gauss mode with respect to a node at $y=0$.
Indeed, our combination of light waves acts as two thin optical lenses covering the domains
$-\infty<y\leq 0$ and $0\leq y<\infty$ and contributing to the total phase 
$\Phi_g$ given by Eq. (\ref{total-phase-Phi}). Depending on the sign of $\Delta$, 
the first lens is converging, whereas the second one is diverging, or vice versa. 
According to Eq. (\ref{focus-optical potential}), the corresponding focal lengths 
\begin{equation}
 \label{focus lengthes}
    {\mathcal F}_{\pm}\equiv 2\kappa v_y^2(\pm\Delta-\omega_{\alpha})=2\kappa v_y^2(\pm\Delta-k_yv_y)
\end{equation}
give rise with the familiar identity 
\begin{equation}
 \label{F formula}
    \frac{1}{{\mathcal F}}=\frac{1}{{\mathcal F}_{+}}+\frac{1}{{\mathcal F}_{-}}
\end{equation}
to the total focal length
\begin{equation}
 \label{focus length qualitative}
    {\mathcal F}=
    2\kappa v_y^2\frac{(\Delta-k_yv_y)(\Delta+k_yv_y)}{2k_yv_y}\approx
    \frac{\Delta}{k_yv_y}{\mathcal F}_0,
\end{equation}
which coincides with Eq. (\ref{focus length}). Here we have recalled the definition 
Eq. (\ref{focus-optical potential}) and used the fact that $k_yv_y\ll|\Delta|$. 
Thus, the suggested atom lens is designed in a similar manner as 
a conventional achromatic lens in optics \cite{MB-EW}. 

The conditions Eqs. (\ref{spontaneous emission}) and (\ref{RN}) 
are satisfied in experiments \cite{Oberthaler-Pfau-Balykin}. 
Indeed, for metastable helium with the velocity $v_y=2000\,{\rm m}/{\rm s}$, 
the angle $\alpha=10^{-3}$, the waist $w_0=50\,{\rm \mu m}$, the wave length $\lambda=1083\,{\rm nm}$, 
the wave packet width $\delta x=\lambda$, the detuning $\Delta=2\pi\times 30\,{\rm MHz}$, 
the rate $\Gamma=10^{7}\,{\rm s}^{-1}$, and the Rabi frequency $\Omega_0=|\Delta|$, 
we obtain the interaction time $2w_0/v_y=50\,{\rm ns}$ and the Doppler frequency $k_yv_y=2\pi\times 1.85\, {\rm MHz}$. 
For these values, we obtain the focal length ${\mathcal F}_0=390\,{\rm \mu m}$, 
the spot size ${\mathcal S}_0=3\,{\rm nm}$, and 
${\mathcal F}_0/{\mathcal F}={\mathcal S}_0/{\mathcal S}=(k_yv_y)/\Delta=0.06$. 

In summary we have proposed a lens with reduced chromatic aberrations.
Our scheme differs from a conventional lens by the use of {\it two} rather than a single light field.
The improvement factor is given by the ratio of the detuning and a Doppler shift.

It is interesting to note that we can interpret \cite{PM}
$\Phi_g$ as a sum of two Berry phases \cite{Berry84,Shapere-Wilczek,Bohm}
acquired by the atom during the two interaction regions $-\infty<y\leq 0$ and $0\leq y<\infty$.
Since the Berry phase is purely of geometrical nature,
it is insensitive to small perturbations in the control parameters \cite{Chiara}.
For this reason, we expect that our so-designed lens is more robust against small fluctuations
of the system parameters, such as the intensity of the light fields.

{\it Acknowledgments.} We are deeply indebted to J. Baudon,
M.V. Fedorov, R. Kaiser, M.K. Oberthaler, R. Walser, and V.P. Yakovlev
for many suggestions and stimulating discussions. MAE is
grateful to the Alexander von Humboldt Stiftung and Russian Foundation for Basic Research (grant 10-02-00914-a).
PVM acknowledges support from the EU project "CONQUEST" and the German Academic
Exchange Service (DAAD).


\begin{thebibliography}{99}
\bibitem{remark} In practice the ''top-hat'' profile is modeled by
a continuous function, for instance, by $\left\{1+\tanh(t/\tau_1)\right\}/2$ with the duration time $\tau_1$.
However, if $\tau_1\ll\tau$, results obtained with the function $f_c(t)$ Eq. (\ref{f1(t)})
are similar to those obtained with the continuous function.
\end{thebibliography}
\end{document}